\documentclass[aps,prx,twocolumn,superscriptaddress,showpacs]{revtex4-1}

\usepackage{amsmath,amssymb,graphics,epsfig,epstopdf,color,verbatim,ulem,braket,tabularx}
\usepackage[colorlinks,linkcolor=blue,citecolor=blue,urlcolor=blue]{hyperref}

\begin{document}

\title{Direct observation of fragile Mott insulators on plaquette Hubbard lattices}

\author{Han-Qing Wu}
\author{Rong-Qiang He}
\affiliation{Department of Physics, Renmin University of China, Beijing 100872, China}
\author{Zi Yang Meng}
\affiliation{Beijing National Laboratory for Condensed Matter Physics,
and Institute of Physics, Chinese Academy of Sciences, Beijing 100190, China}
\author{Zhong-Yi Lu}
\affiliation{Department of Physics, Renmin University of China, Beijing 100872, China}

\email{zlu@ruc.edu.cn}

\begin{abstract}
Employing extensive cellular dynamical mean-field theory (CDMFT) calculations with exact diagonalization impurity solver, we investigate the ground state phase diagrams and non-magnetic metal-insulator transitions of the half-filled Hubbard model on two plaquette -- the 1/5 depleted and checkerboard -- square lattices. We identify three different insulators in the phase diagrams: dimer insulator, antiferromagnetic insulator, and plaquette insulator. And we demonstrate that the plaquette insulator is a novel fragile Mott insulator (FMI) which features a nontrivial one-dimensional irreducible representation of the $C_{4v}$ crystalline point-group and cannot be adiabatically connected to any band insulator with time-reversal symmetry. Furthermore, we study the non-magnetic quantum phase transitions from the metal to the FMI and find that this Mott metal-insulator transition is characterized by the splitting of the non-interacting bands due to interaction effects.
\end{abstract}


\date{\today} \maketitle

\section{INTRODUCTION}

Mott insulators~\cite{Mott1968, Imada1998} are a fundamental phenomenon in strongly correlated quantum many-body physics. At fractional filling (throughout this paper, the ``filling" corresponds to the number of electrons per unit cell and spin projection~\cite{nphys2600}), a material must be metallic according to the conventional band theory. However, it could be a Mott insulator due to dramatic correlation effects. If such a Mott insulator does not break any symmetry and has a spin gap, it will carry fractionalized excitations~\cite{Anderson1987,Kalmeyer1987} and possess a non-trivial topological order~\cite{Kivelson1987,Wen1989,Hastings2004}. At integer filling, a band insulator is likely if there is a gap between the uppermost fully occupied band and the lowermost unoccupied band. However, in the presence of crystalline point-group symmetries, some partially filled bands crossing the Fermi energy may touch with other bands at some high symmetry points in the Brillouin zone. The correlation effects could again forbid a trivial band insulator state; meanwhile other types of symmetric Mott insulating phases, either with~\cite{nphys2600,Roy2012} or without~\cite{PhysRevLett.110.125301, Kimchi2013} topological orders, may emerge. Among the latter cases, one category of novel Mott insulator is dubbed as fragile Mott insulator (FMI)\cite{Yao2007,Yao2010}. A fragile Mott insulator features a nontrivial one-dimensional irreducible representation of crystalline point-group, and cannot be adiabatically connected to any band insulator which respects the time-reversal and the same crystalline point-group symmetry. In this sense, it is the interplay between symmetries of the underlying system and correlation effects that give rise to the fragile Mott insulator phase.

Although proposed in Refs.~\onlinecite{Yao2007,Yao2010}, to the best of our knowledge, there has been no unbiased demonstration of the existence of FMI with advanced numerical approaches in strongly correlated systems. Here, we perform such a systematic study. Employing extensive cellular dynamical mean-field theory (CDMFT)~\cite{Georges1996,Kotliar2006,Maier2005,Kotliar2001}  calculations with exact diagonalization (ED) impurity solver~\cite{Caffarel1994,Liebsch2008,Liebsch2009,Koch2008,RQHe2012}, we investigate the ground state phase diagram and Mott metal-insulator transitions of the half-filled Hubbard model on two plaquette -- the 1/5 depleted and checkerboard -- square lattices. Based on the simulation results and group theory analysis, we unambiguously demonstrate that there exist FMI phases in these systems.

\begin{figure}
  \includegraphics[width=\columnwidth]{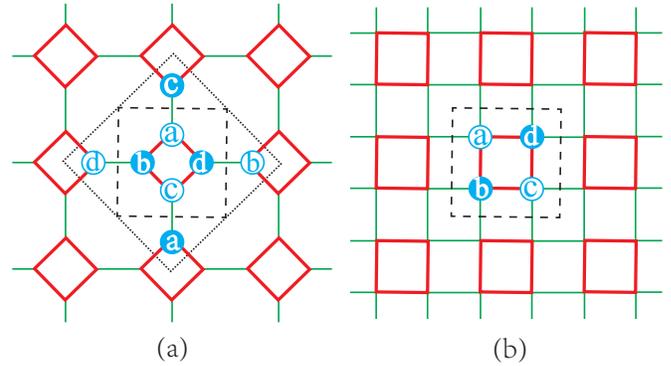}
  \caption{\label{fig:SqLatt}(color online) Illustration of the 1/5-depleted square lattice (a) and checkerboard square lattice (b). Intra (inter)-plaquette hoppings are represented by thick red (thin green) solid lines. The black dashed (dotted) square represents the 4 (8)-site cluster used in our CDMFT+ED calculations, these clusters reflect $C_{4v}$ point-group symmetry. The lowercase letters (a, b, c, d) represent four sites in a unit cell. White dots with blue letters and blue dots with white letters denote the two magnetic sublattices when the system develops antiferromagnetic order.}
\end{figure}

Both 1/5-depleted and checkerboard square lattices consist of coupled plaquette unit cells with four sites per unit cell (see Fig.~\ref{fig:SqLatt}) and are non-Bravais lattices with $C_{4v}$ crystalline point-group symmetry. Here we consider the conventional half-filling case, which corresponds to four electrons within a unit cell and thus belongs to the above defined integer filling case. The 1/5-depleted square lattice~\cite{Satoshi1995,Ueda1996,Troyer1996} was first discovered in the study of spin-gapped calcium vanadate material CaV$_{4}$O$_{9}$~\cite{Satoshi1995}, and later on, in a vacancy-ordered iron selenide family of pnictides~\cite{Bao2011,Ye2011,Yan2011} where a rich variety of phases, including several magnetically ordering and superconducting, have been observed~\cite{Maiti2011,Xu2013}. Recently, the half-filled Hubbard model on this lattice has been studied with different numerical methods, including CDMFT with continuous-time quantum Monte Carlo impurity solver~\cite{Yanagi2014}, determinantal quantum Monte simulations~\cite{Khatami2014}, and variational cluster approximation~\cite{Yamada2014}. However, a systematic study in which magnetic to non-magnetic phase transition, Mott metal-insulator transition, as well as the realization of a fragile Mott insulator phase, has not been carried out. Here, we employ extensive CDMFT+ED simulations to explore the ground state phase diagram and the non-magnetic metal-insulator transitions. We find a fragile Mott insulator phase in this model and confirm its novel symmetry properties and its Mott insulator character, based on numerical results and group theory analysis. In addition, we find that in the checkerboard square lattice \cite{Tsai2006,Wenzel2009,Chakraborty2011,Ying2014}, a fragile Mott insulator also exists. We determine its ground state phase diagram by means of extensive CDMFT+ED simulations as well.

\section{MODEL AND METHOD}

We study a two-dimensional single-band Hubbard model on 1/5-depleted and checkerboard square lattice, as schematically shown in Fig.~\ref{fig:SqLatt}. The Hamiltonian reads,
\begin{eqnarray}
  \hat{H} &=& \hat{H}_{0} + U\sum_{i\alpha}\hat{n}_{i\alpha\uparrow}\hat{n}_{i\alpha\downarrow}, \label{eq:hamiltonian}\\
  \hat{H}_{0} &=& -\sum_{i\alpha,j\beta,\sigma}t_{i\alpha,j\beta}\hat{c}_{i\alpha\sigma}^{\dagger}\hat{c}_{j\beta\sigma} - \mu\sum_{i\alpha\sigma}\hat{n}_{i\alpha\sigma}\nonumber\\
  &=& \sum\limits_{\mathbf{k}\in \text{BZ},\sigma}\hat{\mathbf{c}}_{\mathbf{k}\sigma}^{\dagger}\mathbf{H}_{0}(\mathbf{k})\hat{\mathbf{c}}_{\mathbf{k}\sigma},
\end{eqnarray}
where $\mathbf{H}_{0}(\mathbf{k})$ is the non-interacting Bloch Hamiltonian matrix and $i$ and $j$ label the unit cells, $\alpha$ and $\beta$ label the sites ($a$,$b$,$c$,$d$) within a unit cell. In momentum space, $\hat{\mathbf{c}}_{\mathbf{k}\sigma}^{\dagger}=(\hat{c}_{a\mathbf{k}\sigma}^{\dagger},\hat{c}_{b\mathbf{k}\sigma}^{\dagger},\hat{c}_{c\mathbf{k}\sigma}^{\dagger}, \hat{c}_{d\mathbf{k}\sigma}^{\dagger})$. $U$ is the on-site repulsive Coulomb interaction and we set the chemical potential $\mu=U/2$ for the half-filling. Here $t_{i\alpha,i\beta}=t$ is the intra-plaquette hopping and $t_{i\alpha,j\beta}(i\ne j)=t'$ is the inter-plaquette hopping. As we vary the ratio $t'/t$, the bandwidths of the non-interacting band structures in the two models are fixed at $W=4$. Correspondingly we set the energy unit to be $W/4$ throughout the paper. For simplicity, we introduce a parameter $\lambda\in [0,1]$, specified as follows. For the 1/5-depleted square lattice (Fig.~\ref{fig:SqLatt} (a)):
\vspace{2ex}

  \indent $t=2\lambda/(1+\lambda)$ and $t'=2-4\lambda/(1+\lambda)$;\\
  \indent$\lambda=0$ is the decoupled-dimer limit;\\
  \indent$\lambda=1$ is the decoupled-plaquette limit;\\
  \indent$\lambda=1/2$ is the homogenous case with $t=t'$;\\

\noindent and for the checkerboard square lattice (Fig.~\ref{fig:SqLatt} (b)):
\vspace{2ex}

  \indent $t=\lambda$ and $t'=1-\lambda$;\\
  \indent $\lambda=0$ is one decoupled-plaquette limit;\\
  \indent $\lambda=1$ is another decoupled-plaquette limit;\\
  \indent $\lambda=1/2$ is the homogenous square lattice limit.\\

\begin{figure}
  \includegraphics[width=\columnwidth]{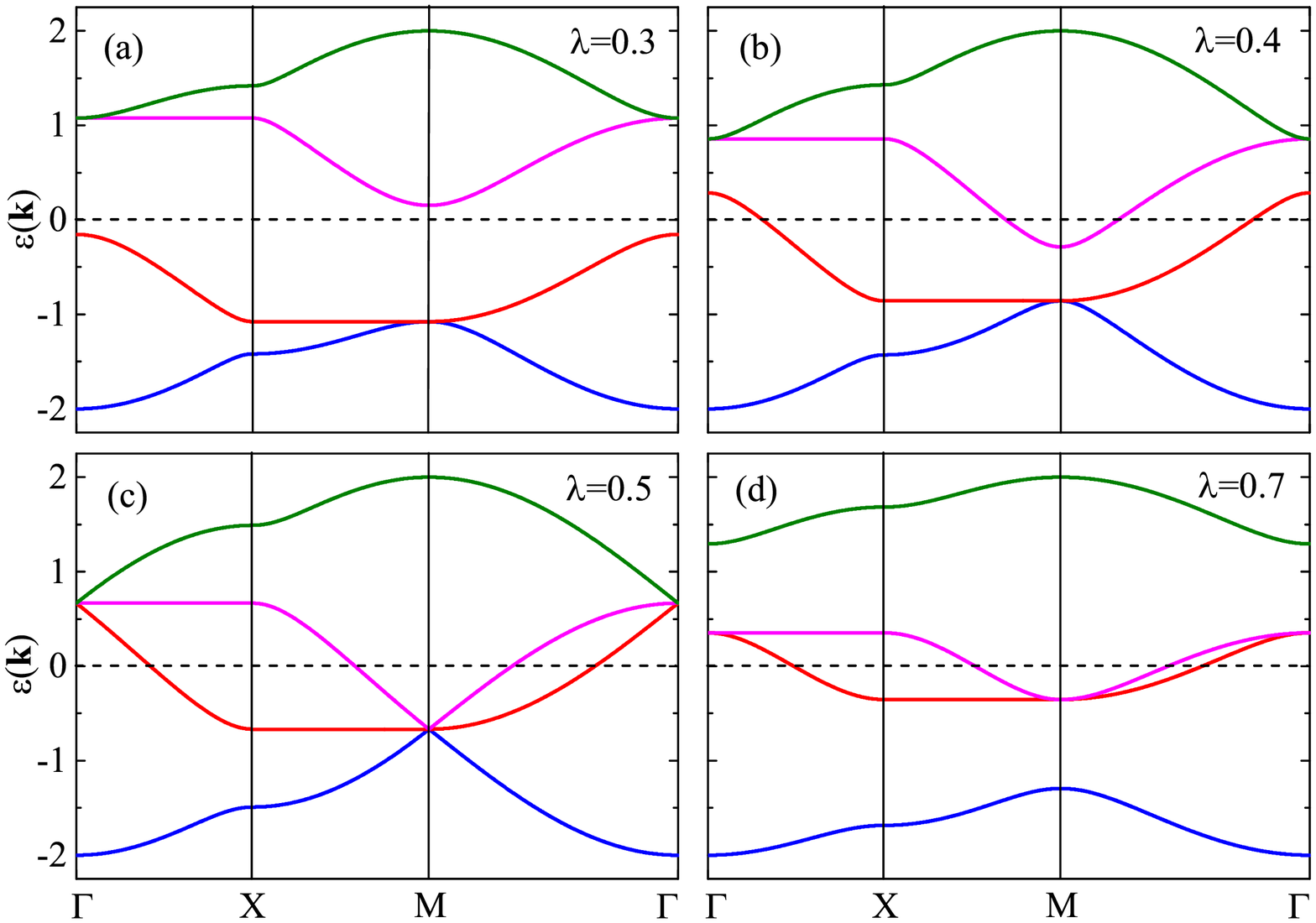}
  \caption{\label{fig:dsband}(color online) Non-interacting band structure of the 1/5-depleted square lattice along the high symmetry path $\Gamma(0,0)\rightarrow\text{X}(0,\pi)\rightarrow\text{M}(\pi,\pi)\rightarrow\Gamma(0,0)$. In the region of $0<\lambda<1/3$, e.g. (a), the system is a band insulator; in the region of $1/3<\lambda<1$, e.g. (b-d), the system becomes a metal with a hole pocket centered at $\Gamma$ point and an electron pocket centered at $M$ point, and the Fermi surface is nested. At $\lambda=1/2$, both $\Gamma$ and M points are threefold degenerated.}
\vspace{2ex}
  \includegraphics[width=\columnwidth]{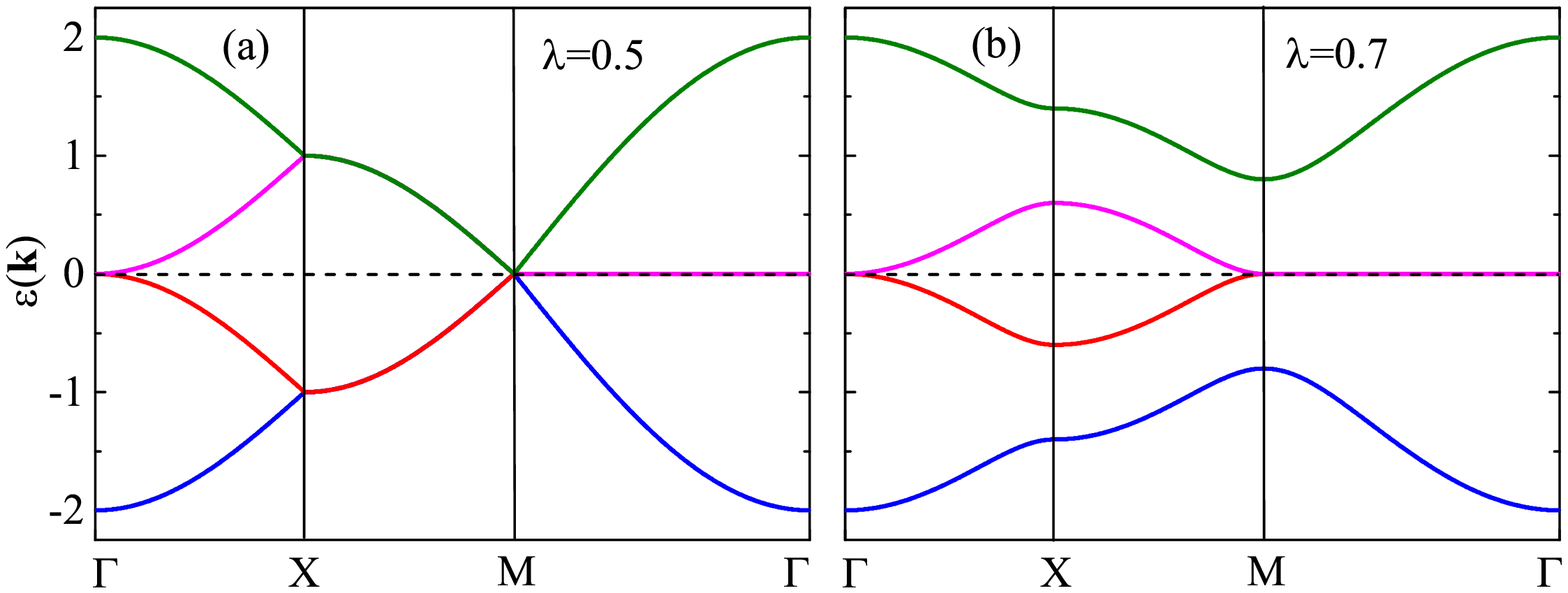}
  \caption{\label{fig:csband}(color online) Non-interacting band structure of the checkerboard square lattice along the high symmetry path. Different from the 1/5-depleted square lattice, there is no band insulator for the whole $\lambda$ range, and the Fermi surface is always nested.}
\end{figure}

Figure \ref{fig:dsband} shows the non-interacting band structure of the 1/5-depleted square lattice. In the half-filling case, there are a band insulating phase in the region of $0 < \lambda < 1/3$ and a metallic phase with nested Fermi surface in the region of $1/3 < \lambda < 1$ \cite{Yanagi2014}, respectively. The non-interacting band structure of the checkerboard square lattice is shown in Fig.~\ref{fig:csband}, in which the system is always metallic with nested Fermi surface for the entire $\lambda$ range.

To study the correlated systems described by Eq.~(\ref{eq:hamiltonian}), we employed CDMFT+ED method. The CDMFT, as a cluster extension of dynamical mean-field theory, maps an interacting lattice problem onto an auxiliary quantum cluster impurity problem embedded in a self-consistently determined mean-field bath. The short-range correlations within the cluster can be treated exactly, while the non-local correlations between clusters are treated at a mean-field level. In this paper, we perform zero temperature CDMFT+ED calculations with 4 (8) correlated impurities (see Fig.~\ref{fig:SqLatt}) in the plaquette (dimer) side of the phase diagram and keep 8 bath levels in total. We introduce the nomenclature ``(ds/cs/s)$n_{c}$-$m_{b}$b-AF/PM/ED"~\cite{Hassan2013} to differentiate the technical details of impurity cluster systems used in the simulations, where ``(ds/cs/s)$n_{c}$-$m_{b}$b" represent the 1/5-depleted/checkerboard/homogenous square lattices with $n_{c}$ correlated impurities and $m_{b}$ bath levels, while ``AF/PM/ED" stand for the CDMFT calculations with the antiferromagnetic mean-field bath, the paramagnetic mean-field bath, or, a purely finite-size Lanczos ED calculation, respectively.

In the CDMFT+ED simulations, the size of impurity system (correlated impurities plus bath levels) cannot be too large as the Hilbert space of the system grows exponentially with its size. However, we have verified that the sizes of the impurity system employed here are sufficient to capture the thermodynamic limit properties of the underlying strongly correlated many-body ground states. Appendix~\ref{app:benchmark} shows our CDMFT+ED simulation results of the impurity system with various sizes on the staggered magnetization for the half-filled Hubbard model on a homogeneous square lattice. The results agree well with the quantum Monte Carlo ones at the Heisenberg limit. For the 1/5-depleted square lattice, we employ the ds8-8b-AF impurity system in the dimer side (labelled by the black dotted diamond in Fig.~\ref{fig:SqLatt} (a)), and this impurity system is able to treat the inter- and intra-plaquette short-range correlations on equal footing. In the plaquette side, however, the ds8-8b-AF is not suitable since it can not correctly take the strong intra-plaquette correlations of the outer four boundary sites in their own plaquettes into consideration. Instead, we use the ds4-8b-AF impurity system (labelled by the black dashed square in Fig.~\ref{fig:SqLatt} (a)) which captures the correlation within a plaquette.

\onecolumngrid

\begin{figure}[h]
  \includegraphics[width=\columnwidth]{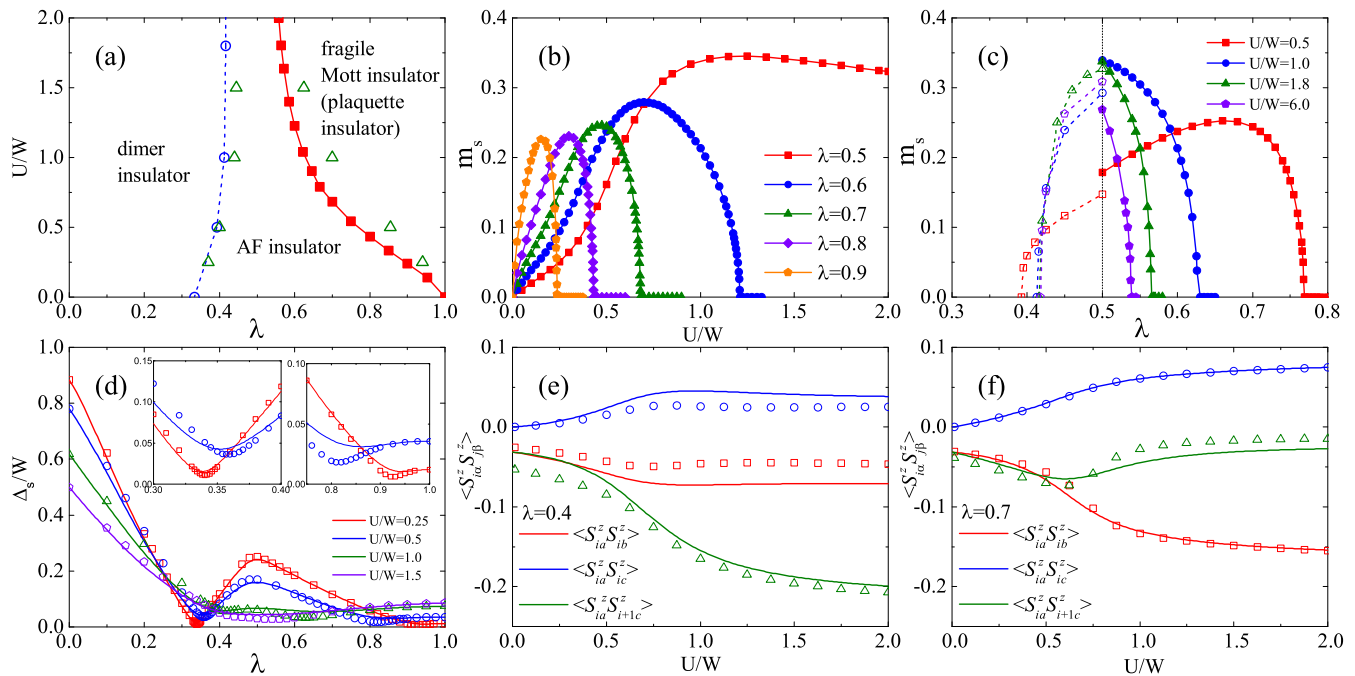}
  \caption{\label{fig:ds_QPT} (color online) (a) Phase diagram of the half-filled Hubbard model on the 1/5-depleted square lattice. The green circles, red squares, and blue triangles are the phase boundaries obtained from ds8-8b-AF, ds4-8b-AF, and ds16-0b-ED systems, respectively. The dimer insulator at small $\lambda$ and the fragile Mott insulator at large $\lambda$ are separated by the AF insulator. The region of AF insulator shrinks as the interaction strength $U/W$ increases. (b-c) Staggered magnetization $m_{s}=\frac{1}{2n_{c}}\sum_{i=1}^{n_{c}}|\braket{\hat{n}_{i\uparrow}}-\braket{\hat{n}_{i\downarrow}}|$ as functions of $\lambda$ and $U/W$, the solid and hollow points are obtained from ds4-8b-AF and ds8-8b-AF impurity systems, respectively. (d) Spin gap $\Delta_{s}=E_{1}(S=1)-E_{0}(S=0)$ are calculated from ds8-0b-ED (solid line) and ds16-0b-ED (hollow points) systems with periodic boundary condition show two minima in the curves which can be used to estimate the quantum phase transitions from dimer insulator to AF insulator and from AF insulator to FMI. Insets are the enlarged plots at the two phase transition regions. (e-f) Spin-spin correlations between intra-plaquette (red, blue) and inter-plaquette (green) sites become more inhomogeneous as $U$ increasing. Solid lines and hollow points are obtained from the ds8-0b-ED and ds16-0b-ED systems, respectively.}
\end{figure}

\twocolumngrid

\section{RESULTS and Analysis}

Figure \ref{fig:ds_QPT}(a) shows the phase diagram of the half-filled Hubbard model on the 1/5-depleted square lattice, obtained from the CDMFT+ED simulations. Three different insulating phases exist: dimer insulator (DI), plaquette insulator (fragile Mott insulator), and the intervening N\'{e}el antiferromagnetic insulator (AFI). The magnetic to non-magnetic phase transitions are continuous, which can be seen from the continuous vanishing of the staggered magnetization $m_{s}$ shown in Fig.~\ref{fig:ds_QPT} (b) and (c). The dimer (plaquette) insulator has all the symmetries of the underlying lattice and is a singlet gapped state which is adiabatically connected to the decoupled dimer (plaquette) limit. In these two spin-gapped phases, the energy gap is a singlet-triplet excitation gap. This gap can be directly calculated by the finite size ED and the corresponding results are shown in Fig.~\ref{fig:ds_QPT} (d). The AFI, with a magnetic long-range order, breaks the SU(2) spin rotational symmetry and then has gapless Goldstone modes. Thus, the gap closing transitions in Fig.~\ref{fig:ds_QPT} (d) signify the DI to AFI and FMI to AFI transitions.

The physical picture of the phase transitions in Fig.~\ref{fig:ds_QPT} (a) can be readily appreciated. The nested Fermi surface in the area of $1/3<\lambda<1$ at $U=0$ (see Fig.~\ref{fig:dsband}) is unstable towards the antiferromagnetic insulating phase upon infinitesimally small $U$. As $U$ is further increased, depending on the value of $\lambda$, the short-range inter-plaquette and intra-plaquette spin correlations start to develop. As can be seen from the spin-spin correlations shown in Fig.~\ref{fig:ds_QPT} (e-f), the system would favor a spin-singlet ground state, either in the form of dimer or plaquette. When the short-range correlation inside the dimer or plaquette is strong enough to destroy the long-range AF order, the continuous quantum phase transitions from AFI to DI or from AFI to FMI occur.

The phase boundaries obtained here are close to those obtained by other methods on the Hubbard model~\cite{Khatami2014, Yamada2014}. Furthermore, extrapolating to the Heisenberg limit, we get the transition points $\lambda_{1c}^{\text{CDMFT}}\approx 0.418$ (obtained by ds8-8-AF) and $\lambda_{2c}^{\text{CDMFT}}\approx 0.536$ (obtained by ds4-8b-AF), which are consistent with the quantum Monte Carlo results for Heisenberg model in Refs.~\onlinecite{Troyer1996,Schwandt2009}, with $\lambda_{1c}^{\text{QMC}}\approx 0.436$ and $\lambda_{2c}^{\text{QMC}}\approx 0.509$, respectively.

Next, we discuss the differences between the dimer insulator and the plaquette insulator, and reveal the fact that the plaquette insulator is essentially a fragile Mott insulator. According to the group theory, a non-degenerate ground state which does not break the crystalline point-group symmetry must transform according to one of the one-dimensional irreducible representations of the crystalline point-group. Therefore, two non-symmetry-breaking phases must be distinct phases, if their respective non-degenerate ground states transform according to different one-dimensional irreducible representations of the point-group. Now, let's consider a finite size system which contains $L\times L$ plaquettes (e.g. the $L=3$ cases in Fig.~\ref{fig:SqLatt}) and preserves the $C_{4v}$ crystalline point-group symmetry. In the decoupled-dimer limit, $\lambda=0$, the non-degenerate ground state of this system is a product state of all the singlets on each inter-plaquette bond. This ground state transforms according to the identity ($A_{1}$) representation of $C_{4v}$ point-group no matter whether $L$ is even or odd. Since the DI phase can be adiabatically connected to the decoupled-dimer limit, the DI phase will transform according to the identity representation. Note, the $U=0$ band insulator with $0<\lambda<1/3$ (Fig.~\ref{fig:ds_QPT} (a)) certainly transforms according to the identity representation~\cite{Yao2010}.

\begin{figure}
  \includegraphics[width=\columnwidth]{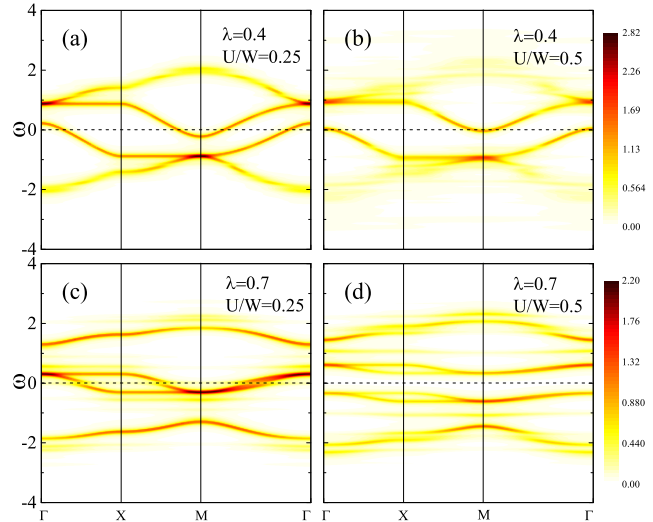}
  \caption{\label{fig:ds_Akw} (color online) Spectral function $A(\mathbf{k},\omega)=-\text{Im}[G(\mathbf{k},\omega+i\eta)]/\pi$ of the half-filled Hubbard model on the 1/5-depleted square lattice along the high symmetry path. Here the Lorentzian broadening factor $\eta$ is $0.05$. (a-b) are obtained from ds8-8b-PM impurity system at $\lambda=0.4$ in the dimer side. We use the self-energy $\Sigma$ periodization scheme~\cite{Sakai2012, Yanagi2014} to restore the unit cell translation symmetry. As $U$ increases, the electron pocket centered at M point and the hole pocket centered at $\Gamma$ point shrink, the system undergoes a Lifshitz transition~\cite{Yanagi2014,Chen2012}. (c-d) Spectral function $A(\mathbf{k},\omega)$ are obtained from ds4-8b-PM impurity system at $\lambda=0.7$ in the plaquette side. The splitting of electronic bands near the Fermi surface is clearly a hallmark of Mott-Hubbard metal-insulator transition, i.e., a direct Mott gap opens by $U$ with the spectral weights transfer to higher energies.}
\end{figure}

The situation is different in the plaquette side. In the decoupled-plaquette limit, $\lambda=1$, the non-degenerate ground state of the $L\times L$ plaquette system is a product of all the singlets on each plaquette. The singlet on a plaquette is an entangled quantum state and has $d_{x^{2}-y^{2}}$ symmetry~\cite{Tsai2006,Yao2007,Yao2010}. This remarkable feature directly determines the nontrivial group symmetry of the plaquette-limit ground state. Following the argument in Ref.~\onlinecite{Yao2010}, we perform detailed group theory analysis, as shown in Appendix~\ref{app:1D}. It follows that the ground state in the plaquette limit transforms according to the nontrivial $B_{2}$ ($A_{1}$) representation when $L$ is odd (even). In the thermodynamic limit, a phase will not depend on the way how the thermodynamic limit is approached, therefore, the plaquette insulator phase, which is continuously connected to the plaquette limit state, cannot be adiabatically connected to DI or any time-reversal symmetric band insulator (as they transform according to the $A_{1}$ representation). Hence, the plaquette insulator is an FMI, as proposed in Refs.~\onlinecite{Yao2007,Yao2010}. Accordingly, there must be some intervening phases or a direct phase transition between the DI and the FMI. And indeed, there is an intervening AFI phase with N\'{e}el antiferromagnetic order between the DI and the FMI in the phase diagram, as shown in Fig.~\ref{fig:ds_QPT} (a).

Furthermore, in the non-interacting $U=0$ limit, there are four bands among which two bands touch at high symmetry points $\Gamma$ and M in the Brillouin zone (see Fig.~\ref{fig:dsband}) due to the wavevector group $G_{\mathbf{k}}=C_{4v}$ at $\mathbf{k}=\text{M}$ or $\Gamma$. The wavevector group is a subgroup of crystalline point-group that leaves the wavevector invariant or translates it by a reciprocal lattice vector. According to the group representation theory, the Bloch Hamiltonian $\mathbf{H}_{0}(\mathbf{k})$ at M or $\Gamma$ point can be block-diagonalized according to the irreducible representations of the wavevector group: $A_{1}\oplus B_{2}\oplus E$. The presence of a two-dimensional irreducible representation $E$ indicates that the two bands are essentially degenerated at $\Gamma$ and M points (see Fig.~\ref{fig:dsband} and Fig.~\ref{fig:ds_Akw}). Taking M point for example, the lower two bands touch and form a two-dimensional irreducible representation($E$) of $G_{\text{M}}$ in the region of $0<\lambda<1/2$, while the middle two bands touch in the region of $1/2<\lambda<1$. The different bands touching at M or $\Gamma$ point have remarkable effects upon the non-magnetic metal-insulator transition, as can be seen from the spectral functions calculated from the CDMFT+ED in Fig.~\ref{fig:ds_Akw}. The non-magnetic metal-insulator transition in the region of $1/3<\lambda<1/2$ is of a Lifshitz type~\cite{Lifshitz1960,Chen2012,Yanagi2014} with shrinking electron and hole pockets and opening an indirect gap at critical $U_{c}$ (see Fig.~\ref{fig:ds_Akw} (a-b)). On the contrary, in the plaquette region of $1/2<\lambda<1$, the correlation effect does not push the middle two quasi-particle bands away from each other, but instead, it splits the bands near the Fermi energy to form an insulator as $U$ increases (see Fig.~\ref{fig:ds_Akw} (c-d)). Thus, the non-magnetic metal-insulator transition in the plaquette side is of a Mott-Hubbard type. In this sense, the plaquette insulator is indeed an FMI because of not only its nontrivial symmetry properties but also being the Mott metal-insulator transition. In addition, the $d$-wave character of the ground state wave function will have some impact on the macroscopic observable properties, such as the $d$-wave symmetry of the pairing correlations~\cite{Yao2007, Khatami2014}. Therefore, we can also call the plaquette insulator a ``$d$-Mott" insulator~\cite{Yao2007, Yao2010}.

\begin{figure}
  \includegraphics[width=\columnwidth]{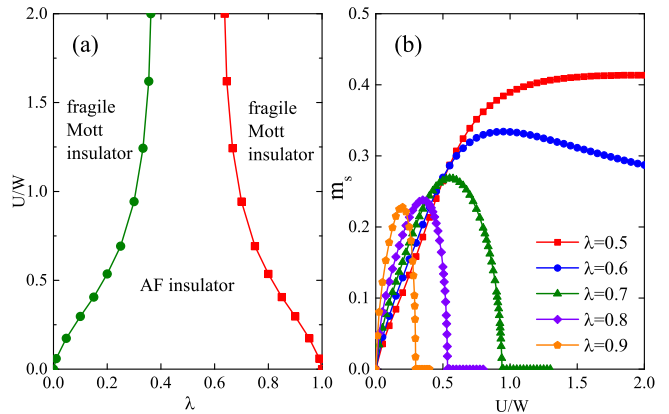}
  \caption{\label{fig:cs_QPT} (color online) Phase diagram and staggered magnetization of the half-filled Hubbard model on the checkerboard square lattice. The results are obtained by the CDMFT+ED with cs4-8b-AF impurity system. We use four-site cluster (see Fig.~\ref{fig:SqLatt} (b)) in the CDMFT simulations in order to accurately incorporate the correlations within plaquette.}
\end{figure}

Now we study the case of the checkerboard square lattice. We obtain two symmetric phase boundaries and further find that the magnetic to non-magnetic phase transitions are also continuous, as shown in Fig.~\ref{fig:cs_QPT}. The extrapolated critical point $\lambda_{2c}^{\text{CDMFT}}\approx 0.62$ is also compatible with quantum Monte Carlo result $\lambda_{2c}^{\text{QMC}}\approx 0.5745$ in the Heisenberg limit~\cite{Lauchli2002,Wenzel2009}. Similarly, the ground states at the two plaquette insulator phases transform according to the nontrivial $B_{1}$ (identity $A_{1}$) representation when $L$ is odd (even). Thus we can use the similar symmetry argument to prove that these two plaquette insulators cannot be adiabatically connected to any time-reversal symmetric band insulator in the thermodynamic limit. The non-interacting band structure in Fig.~\ref{fig:csband} shows that there is no band insulator in the whole $\lambda$ range. The bands touching at $\Gamma$ and M points are also due to the wavevector group $G_{M(\Gamma)}=C_{4v}$. Likewise, the metal-insulator transition is of the Mott-Hubard type. Therefore, the plaquette insulators in this model are also FMI or $d$-Mott insulator, as referred in Refs.~\onlinecite{Tsai2006,Yao2007,Ying2014}.

\section{SUMMARY AND DISCUSSION}

In summary, we have mapped out the ground state phase diagrams of the Hubbard model on the 1/5-depleted and checkerboard square lattices, by means of extensive CDMFT+ED simulations. For the first time being demonstrated with advanced numerical approach, we find out that the plaquette insulators in these systems are actually a well-defined fragile Mott insulator. The FMI is always separated from a band insulator at $U=0$ limit. In the phase diagram of the 1/5-depleted square lattice, an intervening N{\'e}el ordered AFI separates the DI and FMI at finite $U/W$. Our numerical and analytical calculations show that the DI is adiabatically connected to band insulator, while the FMI cannot be adiabatically connected to any time-reversal symmetric band insulator or DI. The non-magnetic metal-insulator transition at the dimer side is of Lifshitz type, while that at the FMI side is of Mott-Hubbard type with splitting of the energy bands crossing the Fermi energy.

As the DI and FMI transform according to different one-dimensional irreducible representations of the $C_{4v}$ point-group symmetry, it will be interesting to investigate possible superconductivity instabilities in the two plaquette systems upon doping. Under the interplay of crystalline point-group symmetry and correlation-driven magnetic properties, a change of the superconducting pairing symmetry is expected~\cite{Scalapino1996,Tsai2006,Khatami2014}. Direct observation of such change of pairing symmetry will shed light on the unconventional superconductivity in the iron selenide family as well as the understanding of the intimate relation between magnetic and superconducting correlations in similar systems.

\begin{acknowledgments}

We thank Hong Yao and Ning-Hua Tong for inspirational discussions. All the numerical calculations were carried out on the Physical Laboratory of High Performance Computing in RUC. This work is supported by National Natural Science Foundation of China (Grant Nos. 91121008 and 11190024) and National Program for Basic Research of MOST of China (Grant No. 2011CBA00112). Z.Y.M. is supported by the National Thousand-Young-Talents Program of China.

\end{acknowledgments}

\appendix

\section{CDMFT+ED benchmark calculations on two-dimensional homogeneous square lattice}
\label{app:benchmark}

\begin{figure}[h]
  \includegraphics[width=\columnwidth]{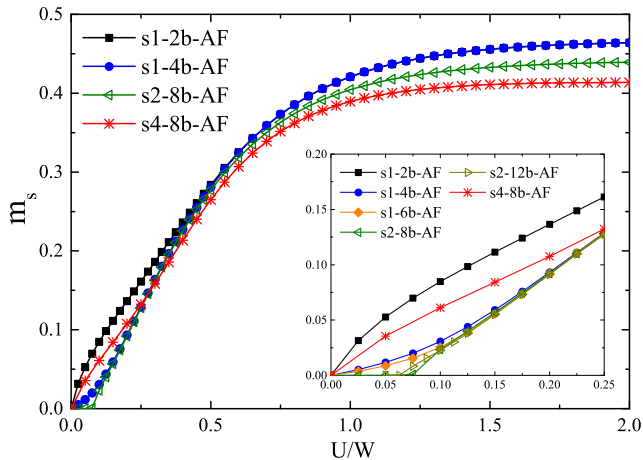}
  \caption{\label{fig:sqlatt_stagmag}(color online) The staggered magnetizations of the Hubbard model on the homogeneous square lattice calculated by CDMFT+ED with several different impurity systems. When $U$ is small, more than two bath levels per cluster boundary site are needed to produce a remarkably good result (see the inset). When $U$ is large, two bath levels per cluster boundary site are sufficient to produce a good result.}
\end{figure}

As a benchmark test, we perform CDMFT+ED simulations for the half-filled Hubbard model on two-dimensional homogeneous square lattice. The results are shown in Fig.~\ref{fig:sqlatt_stagmag}. From the comparison between systems with the same cluster and different number of bath levels, it can be shown that two bath levels per cluster boundary site are sufficient in an insulating regime, but more bath levels are needed to produce a remarkably good result close to $U=0$. Extrapolating to to the Heisenberg limit, the staggered magnetization (not shown in Fig.~\ref{fig:sqlatt_stagmag}) are as follows: 0.422(s1-4b-AF), 0.398(s2-8b-AF), and 0.370(s4-8b-AF). These results do show a remarkable convergence to the quantum Monte Carlo result $m_{s}^{\text{QMC}}=0.3070(3)$~\cite{Sandvik1997} of Heisenberg model on the homogeneous square lattice.

For the lattice models we studied in this paper, the phase boundaries lie between two insulating phases, see Fig. \ref{fig:ds_QPT} (a) and Fig. \ref{fig:cs_QPT} (a). In addition, the nonlocal correlations are short-range in the plaquette insulator and dimer insulator regime. Therefore, although we use small clusters and total 8 bath levels in the CDMFT+ED calculations, we have good reason to expect that our numerical results can qualitatively well describe magnetic phase transitions and the insulating ground states.

\section{one-dimensional irreducible representation of decoupled plaquette state}
\label{app:1D}

The two generators of the $C_{4v}$ group are the clockwise $\pi/4$ rotation about the $\mathit{z}$ axis, $c_{4}$, and the reflection in a vertical plane, $\sigma_{v}$. The group character table is shown in Table I. There are four different one-dimensional irreducible representations $(A_{1},A_{2},B_{1},B_{2})$ and one two-dimensional irreducible representation $(E)$. A unique ground state which preserves the $C_{4v}$ symmetry must transform according to one of the one-dimensional irreducible representations.

\begin{table}[h]
\begin{ruledtabular}
\begin{tabular}{c | c c c c c}
$C_{4v}$   & $E$   & $c_{2}$  & 2$c_{4}$   & 2$\sigma_{v}$   & 2$\sigma_{d}$ \\
\hline
$A_{1}$    & 1      &  1      &  1      &  1      &  1  \\
$A_{2}$    & 1      &  1      &  1      & -1      & -1  \\
$B_{1}$    & 1      &  1      & -1      &  1      & -1  \\
$B_{2}$    & 1      &  1      & -1      & -1      &  1  \\
$E$        & 2      & -2      &  0      &  0      &  0  \\
\end{tabular}
\label{Table:C4v}
\caption{The character table of $C_{4v}$ point-group.}
\end{ruledtabular}
\end{table}

To simplify the following discussion about the decoupled plaquette state at $\lambda=1$, we can go to the Heisenberg limit. The unique many-body ground state wave function can be taken as the direct product of singlet states on each plaquette,
\begin{equation}
\ket{\Psi_{\lambda=1}^{L}}=\prod_{n=1}^{L\times L}\hat{d}_{n}^{\dagger}\ket{0}
\end{equation}
where $\hat{d}_{n}^{\dagger}=(\hat{s}_{n,ab}^{\dagger}\hat{s}_{n,cd}^{\dagger}-\hat{s}_{n,ad}^{\dagger}\hat{s}_{n,bc}^{\dagger})/\sqrt{3}$ and $\hat{s}_{n,ab}^{\dagger}=(\hat{c}_{n,a\uparrow}^{\dagger}\hat{c}_{n,b\downarrow}^{\dagger}-\hat{c}_{n,a\downarrow}^{\dagger}\hat{c}_{n,b\uparrow}^{\dagger})/\sqrt{2}$. $\hat{d}_{n}^{\dagger}\ket{0}$ creates an plaquette singlet state at the $n$-th plaquette on the $L\times L$ plaquette lattice, and $\hat{s}_{n,ab}^{\dagger}\ket{0}$ creates an $ab$ bond singlet within the $n$-th plaquette (see Fig.~\ref{fig:SqLatt}). Applying the $c_{4}$ symmetry operation,
\begin{equation}
\begin{split}
\hat{P}_{c_{4}}\hat{d}_{n}^{\dagger}\hat{P}_{c_{4}}^{-1}&=\frac{1}{\sqrt{3}}
(\hat{s}_{m,bc}^{\dagger}\hat{s}_{m,da}^{\dagger}-\hat{s}_{m,ba}^{\dagger}\hat{s}_{m,cd}^{\dagger}) \\
&=-\frac{1}{\sqrt{3}}
(\hat{s}_{m,ab}^{\dagger}\hat{s}_{m,cd}^{\dagger}-\hat{s}_{m,ad}^{\dagger}\hat{s}_{m,bc}^{\dagger}) \\
&=-\hat{d}_{m}^{\dagger}
\end{split}
\end{equation}
where $\hat{P}_{c_{4}}$ is a symmetry operator of the corresponding group element $c_{4}$. We can demonstrate that the ground state transforms as
\begin{equation}
\begin{split}
&\hat{P}_{c_{4}}\ket{\Psi_{\lambda=1}^{L}} \\
&=\hat{P}_{c_{4}}\hat{d}_{1}^{\dagger}
\hat{P}_{c_{4}}^{-1}\hat{P}_{c_{4}}\hat{d}_{2}^{\dagger}\cdots\hat{P}_{c_{4}}^{-1}
\hat{d}_{L\times L}^{\dagger}\hat{P}_{c_{4}}\hat{P}_{c_{4}}^{-1}\ket{0} \\
&=(-1)^{L\times L}\hat{d}_{i_{1}}^{\dagger}\hat{d}_{i_{2}}^{\dagger}\cdots\hat{d}_{i_{L\times L}}^{\dagger}\ket{0}\\
&=\left\{
    \begin{array}{ll}
      -\ket{\Psi_{\lambda=1}^{L}}, & \hbox{$L$ odd;} \\
      +\ket{\Psi_{\lambda=1}^{L}}, & \hbox{$L$ even.}
    \end{array}
  \right.
\end{split}
\end{equation}
where we have used the commutation relation $[\hat{d}_{n}^{\dagger}, \hat{d}_{m}^{\dagger}]=0$. This result holds for both the 1/5-depleted and checkerboard square lattices.

In a similar way, we can prove the ground state transforms as
\begin{equation}
\hat{P}_{\sigma_{v}}\ket{\Psi_{\lambda=1}^{L}}
=\left\{
   \begin{array}{ll}
     -\ket{\Psi_{\lambda=1}^{L}}, & \hbox{$L$ odd;} \\
     +\ket{\Psi_{\lambda=1}^{L}}, & \hbox{$L$ even.}
   \end{array}
 \right.
\end{equation}
for the 1/5-depleted square lattice and
\begin{equation}
\hat{P}_{\sigma_{v}}\ket{\Psi_{\lambda=1}^{L}}
=\left\{
   \begin{array}{ll}
     +\ket{\Psi_{\lambda=1}^{L}}, & \hbox{$L$ odd;} \\
     +\ket{\Psi_{\lambda=1}^{L}}, & \hbox{$L$ even.}
   \end{array}
 \right.
\end{equation}
for the checkerboard square lattice. Other eigenvalues of $C_{4v}$ group operators can also be deduced. Together with Table I, we can show that the ground state of the 1/5-depleted (checkerboard) square lattice at the decoupled plaquette limit transforms according to the nontrivial $B_{2}$ ($B_{1}$) one-dimensional irreducible representation when $L$ is odd.

\bibliography{FMI_CDMFT}

\begin{thebibliography}{45}%
\makeatletter
\providecommand \@ifxundefined [1]{%
 \@ifx{#1\undefined}
}%
\providecommand \@ifnum [1]{%
 \ifnum #1\expandafter \@firstoftwo
 \else \expandafter \@secondoftwo
 \fi
}%
\providecommand \@ifx [1]{%
 \ifx #1\expandafter \@firstoftwo
 \else \expandafter \@secondoftwo
 \fi
}%
\providecommand \natexlab [1]{#1}%
\providecommand \enquote  [1]{``#1''}%
\providecommand \bibnamefont  [1]{#1}%
\providecommand \bibfnamefont [1]{#1}%
\providecommand \citenamefont [1]{#1}%
\providecommand \href@noop [0]{\@secondoftwo}%
\providecommand \href [0]{\begingroup \@sanitize@url \@href}%
\providecommand \@href[1]{\@@startlink{#1}\@@href}%
\providecommand \@@href[1]{\endgroup#1\@@endlink}%
\providecommand \@sanitize@url [0]{\catcode `\\12\catcode `\$12\catcode
  `\&12\catcode `\#12\catcode `\^12\catcode `\_12\catcode `\%12\relax}%
\providecommand \@@startlink[1]{}%
\providecommand \@@endlink[0]{}%
\providecommand \url  [0]{\begingroup\@sanitize@url \@url }%
\providecommand \@url [1]{\endgroup\@href {#1}{\urlprefix }}%
\providecommand \urlprefix  [0]{URL }%
\providecommand \Eprint [0]{\href }%
\providecommand \doibase [0]{http://dx.doi.org/}%
\providecommand \selectlanguage [0]{\@gobble}%
\providecommand \bibinfo  [0]{\@secondoftwo}%
\providecommand \bibfield  [0]{\@secondoftwo}%
\providecommand \translation [1]{[#1]}%
\providecommand \BibitemOpen [0]{}%
\providecommand \bibitemStop [0]{}%
\providecommand \bibitemNoStop [0]{.\EOS\space}%
\providecommand \EOS [0]{\spacefactor3000\relax}%
\providecommand \BibitemShut  [1]{\csname bibitem#1\endcsname}%
\let\auto@bib@innerbib\@empty
\bibitem [{\citenamefont {Mott}(1968)}]{Mott1968}%
  \BibitemOpen
  \bibfield  {author} {\bibinfo {author} {\bibfnamefont {N.~F.}\ \bibnamefont
  {Mott}},\ }\href {\doibase 10.1103/RevModPhys.40.677} {\bibfield  {journal}
  {\bibinfo  {journal} {Rev. Mod. Phys.}\ }\textbf {\bibinfo {volume} {40}},\
  \bibinfo {pages} {677} (\bibinfo {year} {1968})}\BibitemShut {NoStop}%
\bibitem [{\citenamefont {Imada}\ \emph {et~al.}(1998)\citenamefont {Imada},
  \citenamefont {Fujimori},\ and\ \citenamefont {Tokura}}]{Imada1998}%
  \BibitemOpen
  \bibfield  {author} {\bibinfo {author} {\bibfnamefont {M.}~\bibnamefont
  {Imada}}, \bibinfo {author} {\bibfnamefont {A.}~\bibnamefont {Fujimori}}, \
  and\ \bibinfo {author} {\bibfnamefont {Y.}~\bibnamefont {Tokura}},\ }\href
  {\doibase 10.1103/RevModPhys.70.1039} {\bibfield  {journal} {\bibinfo
  {journal} {Rev. Mod. Phys.}\ }\textbf {\bibinfo {volume} {70}},\ \bibinfo
  {pages} {1039} (\bibinfo {year} {1998})}\BibitemShut {NoStop}%
\bibitem [{\citenamefont {Parameswaran}\ \emph
  {et~al.}(2013{\natexlab{a}})\citenamefont {Parameswaran}, \citenamefont
  {Turner}, \citenamefont {Arovas},\ and\ \citenamefont
  {Vishwanath}}]{nphys2600}%
  \BibitemOpen
  \bibfield  {author} {\bibinfo {author} {\bibfnamefont {S.~A.}\ \bibnamefont
  {Parameswaran}}, \bibinfo {author} {\bibfnamefont {A.~M.}\ \bibnamefont
  {Turner}}, \bibinfo {author} {\bibfnamefont {D.~P.}\ \bibnamefont {Arovas}},
  \ and\ \bibinfo {author} {\bibfnamefont {A.}~\bibnamefont {Vishwanath}},\
  }\href {\doibase 10.1038/nphys2600} {\bibfield  {journal} {\bibinfo
  {journal} {Nat. Phys.}\ }\textbf {\bibinfo {volume} {9}},\ \bibinfo {pages}
  {299} (\bibinfo {year} {2013}{\natexlab{a}})}\BibitemShut {NoStop}%
\bibitem [{\citenamefont {ANDERSON}(1987)}]{Anderson1987}%
  \BibitemOpen
  \bibfield  {author} {\bibinfo {author} {\bibfnamefont {P.~W.}\ \bibnamefont
  {ANDERSON}},\ }\href {\doibase 10.1126/science.235.4793.1196} {\bibfield
  {journal} {\bibinfo  {journal} {Science}\ }\textbf {\bibinfo {volume}
  {235}},\ \bibinfo {pages} {1196} (\bibinfo {year} {1987})}\BibitemShut
  {NoStop}%
\bibitem [{\citenamefont {Kalmeyer}\ and\ \citenamefont
  {Laughlin}(1987)}]{Kalmeyer1987}%
  \BibitemOpen
  \bibfield  {author} {\bibinfo {author} {\bibfnamefont {V.}~\bibnamefont
  {Kalmeyer}}\ and\ \bibinfo {author} {\bibfnamefont {R.~B.}\ \bibnamefont
  {Laughlin}},\ }\href {\doibase 10.1103/PhysRevLett.59.2095} {\bibfield
  {journal} {\bibinfo  {journal} {Phys. Rev. Lett.}\ }\textbf {\bibinfo
  {volume} {59}},\ \bibinfo {pages} {2095} (\bibinfo {year}
  {1987})}\BibitemShut {NoStop}%
\bibitem [{\citenamefont {Kivelson}\ \emph {et~al.}(1987)\citenamefont
  {Kivelson}, \citenamefont {Rokhsar},\ and\ \citenamefont
  {Sethna}}]{Kivelson1987}%
  \BibitemOpen
  \bibfield  {author} {\bibinfo {author} {\bibfnamefont {S.~A.}\ \bibnamefont
  {Kivelson}}, \bibinfo {author} {\bibfnamefont {D.~S.}\ \bibnamefont
  {Rokhsar}}, \ and\ \bibinfo {author} {\bibfnamefont {J.~P.}\ \bibnamefont
  {Sethna}},\ }\href {\doibase 10.1103/PhysRevB.35.8865} {\bibfield  {journal}
  {\bibinfo  {journal} {Phys. Rev. B}\ }\textbf {\bibinfo {volume} {35}},\
  \bibinfo {pages} {8865} (\bibinfo {year} {1987})}\BibitemShut {NoStop}%
\bibitem [{\citenamefont {Wen}(1989)}]{Wen1989}%
  \BibitemOpen
  \bibfield  {author} {\bibinfo {author} {\bibfnamefont {X.~G.}\ \bibnamefont
  {Wen}},\ }\href {\doibase 10.1103/PhysRevB.40.7387} {\bibfield  {journal}
  {\bibinfo  {journal} {Phys. Rev. B}\ }\textbf {\bibinfo {volume} {40}},\
  \bibinfo {pages} {7387} (\bibinfo {year} {1989})}\BibitemShut {NoStop}%
\bibitem [{\citenamefont {Hastings}(2004)}]{Hastings2004}%
  \BibitemOpen
  \bibfield  {author} {\bibinfo {author} {\bibfnamefont {M.~B.}\ \bibnamefont
  {Hastings}},\ }\href {\doibase 10.1103/PhysRevB.69.104431} {\bibfield
  {journal} {\bibinfo  {journal} {Phys. Rev. B}\ }\textbf {\bibinfo {volume}
  {69}},\ \bibinfo {pages} {104431} (\bibinfo {year} {2004})}\BibitemShut
  {NoStop}%
\bibitem [{\citenamefont {Roy}(2012)}]{Roy2012}%
  \BibitemOpen
  \bibfield  {author} {\bibinfo {author} {\bibfnamefont {R.}~\bibnamefont
  {Roy}},\ }\href {\doibase arXiv:1212.2944} {\bibfield  {journal} {\bibinfo
  {journal} {arXiv}\ }\textbf {\bibinfo {volume} {1212}},\ \bibinfo {pages}
  {2944} (\bibinfo {year} {2012})}\BibitemShut {NoStop}%
\bibitem [{\citenamefont {Parameswaran}\ \emph
  {et~al.}(2013{\natexlab{b}})\citenamefont {Parameswaran}, \citenamefont
  {Kimchi}, \citenamefont {Turner}, \citenamefont {Stamper-Kurn},\ and\
  \citenamefont {Vishwanath}}]{PhysRevLett.110.125301}%
  \BibitemOpen
  \bibfield  {author} {\bibinfo {author} {\bibfnamefont {S.~A.}\ \bibnamefont
  {Parameswaran}}, \bibinfo {author} {\bibfnamefont {I.}~\bibnamefont
  {Kimchi}}, \bibinfo {author} {\bibfnamefont {A.~M.}\ \bibnamefont {Turner}},
  \bibinfo {author} {\bibfnamefont {D.~M.}\ \bibnamefont {Stamper-Kurn}}, \
  and\ \bibinfo {author} {\bibfnamefont {A.}~\bibnamefont {Vishwanath}},\
  }\href {\doibase 10.1103/PhysRevLett.110.125301} {\bibfield  {journal}
  {\bibinfo  {journal} {Phys. Rev. Lett.}\ }\textbf {\bibinfo {volume} {110}},\
  \bibinfo {pages} {125301} (\bibinfo {year} {2013}{\natexlab{b}})}\BibitemShut
  {NoStop}%
\bibitem [{\citenamefont {Kimchi}\ \emph {et~al.}(2013)\citenamefont {Kimchi},
  \citenamefont {Parameswaran}, \citenamefont {Turner}, \citenamefont {Wang},\
  and\ \citenamefont {Vishwanath}}]{Kimchi2013}%
  \BibitemOpen
  \bibfield  {author} {\bibinfo {author} {\bibfnamefont {I.}~\bibnamefont
  {Kimchi}}, \bibinfo {author} {\bibfnamefont {S.~A.}\ \bibnamefont
  {Parameswaran}}, \bibinfo {author} {\bibfnamefont {A.~M.}\ \bibnamefont
  {Turner}}, \bibinfo {author} {\bibfnamefont {F.}~\bibnamefont {Wang}}, \ and\
  \bibinfo {author} {\bibfnamefont {A.}~\bibnamefont {Vishwanath}},\ }\href
  {\doibase 10.1073/pnas.1307245110} {\bibfield  {journal} {\bibinfo  {journal}
  {Proc. Natl. Acad. Sci. U.S.A.}\ }\textbf {\bibinfo {volume} {110}},\
  \bibinfo {pages} {16378} (\bibinfo {year} {2013})}\BibitemShut {NoStop}%
\bibitem [{\citenamefont {Yao}\ \emph {et~al.}(2007)\citenamefont {Yao},
  \citenamefont {Tsai},\ and\ \citenamefont {Kivelson}}]{Yao2007}%
  \BibitemOpen
  \bibfield  {author} {\bibinfo {author} {\bibfnamefont {H.}~\bibnamefont
  {Yao}}, \bibinfo {author} {\bibfnamefont {W.-F.}\ \bibnamefont {Tsai}}, \
  and\ \bibinfo {author} {\bibfnamefont {S.~A.}\ \bibnamefont {Kivelson}},\
  }\href {\doibase 10.1103/PhysRevB.76.161104} {\bibfield  {journal} {\bibinfo
  {journal} {Phys. Rev. B}\ }\textbf {\bibinfo {volume} {76}},\ \bibinfo
  {pages} {161104} (\bibinfo {year} {2007})}\BibitemShut {NoStop}%
\bibitem [{\citenamefont {Yao}\ and\ \citenamefont {Kivelson}(2010)}]{Yao2010}%
  \BibitemOpen
  \bibfield  {author} {\bibinfo {author} {\bibfnamefont {H.}~\bibnamefont
  {Yao}}\ and\ \bibinfo {author} {\bibfnamefont {S.~A.}\ \bibnamefont
  {Kivelson}},\ }\href {\doibase 10.1103/PhysRevLett.105.166402} {\bibfield
  {journal} {\bibinfo  {journal} {Phys. Rev. Lett.}\ }\textbf {\bibinfo
  {volume} {105}},\ \bibinfo {pages} {166402} (\bibinfo {year}
  {2010})}\BibitemShut {NoStop}%
\bibitem [{\citenamefont {Georges}\ \emph {et~al.}(1996)\citenamefont
  {Georges}, \citenamefont {Kotliar}, \citenamefont {Krauth},\ and\
  \citenamefont {Rozenberg}}]{Georges1996}%
  \BibitemOpen
  \bibfield  {author} {\bibinfo {author} {\bibfnamefont {A.}~\bibnamefont
  {Georges}}, \bibinfo {author} {\bibfnamefont {G.}~\bibnamefont {Kotliar}},
  \bibinfo {author} {\bibfnamefont {W.}~\bibnamefont {Krauth}}, \ and\ \bibinfo
  {author} {\bibfnamefont {M.~J.}\ \bibnamefont {Rozenberg}},\ }\href {\doibase
  10.1103/RevModPhys.68.13} {\bibfield  {journal} {\bibinfo  {journal} {Rev.
  Mod. Phys.}\ }\textbf {\bibinfo {volume} {68}},\ \bibinfo {pages} {13}
  (\bibinfo {year} {1996})}\BibitemShut {NoStop}%
\bibitem [{\citenamefont {Kotliar}\ \emph {et~al.}(2006)\citenamefont
  {Kotliar}, \citenamefont {Savrasov}, \citenamefont {Haule}, \citenamefont
  {Oudovenko}, \citenamefont {Parcollet},\ and\ \citenamefont
  {Marianetti}}]{Kotliar2006}%
  \BibitemOpen
  \bibfield  {author} {\bibinfo {author} {\bibfnamefont {G.}~\bibnamefont
  {Kotliar}}, \bibinfo {author} {\bibfnamefont {S.~Y.}\ \bibnamefont
  {Savrasov}}, \bibinfo {author} {\bibfnamefont {K.}~\bibnamefont {Haule}},
  \bibinfo {author} {\bibfnamefont {V.~S.}\ \bibnamefont {Oudovenko}}, \bibinfo
  {author} {\bibfnamefont {O.}~\bibnamefont {Parcollet}}, \ and\ \bibinfo
  {author} {\bibfnamefont {C.~A.}\ \bibnamefont {Marianetti}},\ }\href
  {\doibase 10.1103/RevModPhys.78.865} {\bibfield  {journal} {\bibinfo
  {journal} {Rev. Mod. Phys.}\ }\textbf {\bibinfo {volume} {78}},\ \bibinfo
  {pages} {865} (\bibinfo {year} {2006})}\BibitemShut {NoStop}%
\bibitem [{\citenamefont {Maier}\ \emph {et~al.}(2005)\citenamefont {Maier},
  \citenamefont {Jarrell}, \citenamefont {Pruschke},\ and\ \citenamefont
  {Hettler}}]{Maier2005}%
  \BibitemOpen
  \bibfield  {author} {\bibinfo {author} {\bibfnamefont {T.}~\bibnamefont
  {Maier}}, \bibinfo {author} {\bibfnamefont {M.}~\bibnamefont {Jarrell}},
  \bibinfo {author} {\bibfnamefont {T.}~\bibnamefont {Pruschke}}, \ and\
  \bibinfo {author} {\bibfnamefont {M.~H.}\ \bibnamefont {Hettler}},\ }\href
  {\doibase 10.1103/RevModPhys.77.1027} {\bibfield  {journal} {\bibinfo
  {journal} {Rev. Mod. Phys.}\ }\textbf {\bibinfo {volume} {77}},\ \bibinfo
  {pages} {1027} (\bibinfo {year} {2005})}\BibitemShut {NoStop}%
\bibitem [{\citenamefont {Kotliar}\ \emph {et~al.}(2001)\citenamefont
  {Kotliar}, \citenamefont {Savrasov}, \citenamefont {P\'alsson},\ and\
  \citenamefont {Biroli}}]{Kotliar2001}%
  \BibitemOpen
  \bibfield  {author} {\bibinfo {author} {\bibfnamefont {G.}~\bibnamefont
  {Kotliar}}, \bibinfo {author} {\bibfnamefont {S.~Y.}\ \bibnamefont
  {Savrasov}}, \bibinfo {author} {\bibfnamefont {G.}~\bibnamefont {P\'alsson}},
  \ and\ \bibinfo {author} {\bibfnamefont {G.}~\bibnamefont {Biroli}},\ }\href
  {\doibase 10.1103/PhysRevLett.87.186401} {\bibfield  {journal} {\bibinfo
  {journal} {Phys. Rev. Lett.}\ }\textbf {\bibinfo {volume} {87}},\ \bibinfo
  {pages} {186401} (\bibinfo {year} {2001})}\BibitemShut {NoStop}%
\bibitem [{\citenamefont {Caffarel}\ and\ \citenamefont
  {Krauth}(1994)}]{Caffarel1994}%
  \BibitemOpen
  \bibfield  {author} {\bibinfo {author} {\bibfnamefont {M.}~\bibnamefont
  {Caffarel}}\ and\ \bibinfo {author} {\bibfnamefont {W.}~\bibnamefont
  {Krauth}},\ }\href {\doibase 10.1103/PhysRevLett.72.1545} {\bibfield
  {journal} {\bibinfo  {journal} {Phys. Rev. Lett.}\ }\textbf {\bibinfo
  {volume} {72}},\ \bibinfo {pages} {1545} (\bibinfo {year}
  {1994})}\BibitemShut {NoStop}%
\bibitem [{\citenamefont {Liebsch}\ \emph {et~al.}(2008)\citenamefont
  {Liebsch}, \citenamefont {Ishida},\ and\ \citenamefont
  {Merino}}]{Liebsch2008}%
  \BibitemOpen
  \bibfield  {author} {\bibinfo {author} {\bibfnamefont {A.}~\bibnamefont
  {Liebsch}}, \bibinfo {author} {\bibfnamefont {H.}~\bibnamefont {Ishida}}, \
  and\ \bibinfo {author} {\bibfnamefont {J.}~\bibnamefont {Merino}},\ }\href
  {\doibase 10.1103/PhysRevB.78.165123} {\bibfield  {journal} {\bibinfo
  {journal} {Phys. Rev. B}\ }\textbf {\bibinfo {volume} {78}},\ \bibinfo
  {pages} {165123} (\bibinfo {year} {2008})}\BibitemShut {NoStop}%
\bibitem [{\citenamefont {Liebsch}\ and\ \citenamefont
  {Tong}(2009)}]{Liebsch2009}%
  \BibitemOpen
  \bibfield  {author} {\bibinfo {author} {\bibfnamefont {A.}~\bibnamefont
  {Liebsch}}\ and\ \bibinfo {author} {\bibfnamefont {N.-H.}\ \bibnamefont
  {Tong}},\ }\href {\doibase 10.1103/PhysRevB.80.165126} {\bibfield  {journal}
  {\bibinfo  {journal} {Phys. Rev. B}\ }\textbf {\bibinfo {volume} {80}},\
  \bibinfo {pages} {165126} (\bibinfo {year} {2009})}\BibitemShut {NoStop}%
\bibitem [{\citenamefont {Koch}\ \emph {et~al.}(2008)\citenamefont {Koch},
  \citenamefont {Sangiovanni},\ and\ \citenamefont {Gunnarsson}}]{Koch2008}%
  \BibitemOpen
  \bibfield  {author} {\bibinfo {author} {\bibfnamefont {E.}~\bibnamefont
  {Koch}}, \bibinfo {author} {\bibfnamefont {G.}~\bibnamefont {Sangiovanni}}, \
  and\ \bibinfo {author} {\bibfnamefont {O.}~\bibnamefont {Gunnarsson}},\
  }\href {\doibase 10.1103/PhysRevB.78.115102} {\bibfield  {journal} {\bibinfo
  {journal} {Phys. Rev. B}\ }\textbf {\bibinfo {volume} {78}},\ \bibinfo
  {pages} {115102} (\bibinfo {year} {2008})}\BibitemShut {NoStop}%
\bibitem [{\citenamefont {He}\ and\ \citenamefont {Lu}(2012)}]{RQHe2012}%
  \BibitemOpen
  \bibfield  {author} {\bibinfo {author} {\bibfnamefont {R.-Q.}\ \bibnamefont
  {He}}\ and\ \bibinfo {author} {\bibfnamefont {Z.-Y.}\ \bibnamefont {Lu}},\
  }\href {\doibase 10.1103/PhysRevB.86.045105} {\bibfield  {journal} {\bibinfo
  {journal} {Phys. Rev. B}\ }\textbf {\bibinfo {volume} {86}},\ \bibinfo
  {pages} {045105} (\bibinfo {year} {2012})}\BibitemShut {NoStop}%
\bibitem [{\citenamefont {Satoshi}\ \emph {et~al.}(1995)\citenamefont
  {Satoshi}, \citenamefont {Takashi}, \citenamefont {Yukio}, \citenamefont
  {Yoshiaki}, \citenamefont {Masatoshi}, \citenamefont {Takashi}, \citenamefont
  {Masaaki},\ and\ \citenamefont {Kazuhiro}}]{Satoshi1995}%
  \BibitemOpen
  \bibfield  {author} {\bibinfo {author} {\bibfnamefont {T.}~\bibnamefont
  {Satoshi}}, \bibinfo {author} {\bibfnamefont {N.}~\bibnamefont {Takashi}},
  \bibinfo {author} {\bibfnamefont {Y.}~\bibnamefont {Yukio}}, \bibinfo
  {author} {\bibfnamefont {K.}~\bibnamefont {Yoshiaki}}, \bibinfo {author}
  {\bibfnamefont {S.}~\bibnamefont {Masatoshi}}, \bibinfo {author}
  {\bibfnamefont {N.}~\bibnamefont {Takashi}}, \bibinfo {author} {\bibfnamefont
  {K.}~\bibnamefont {Masaaki}}, \ and\ \bibinfo {author} {\bibfnamefont
  {S.}~\bibnamefont {Kazuhiro}},\ }\href {\doibase 10.1143/JPSJ.64.2758}
  {\bibfield  {journal} {\bibinfo  {journal} {J. Phys. Soc. Jpn.}\ }\textbf
  {\bibinfo {volume} {64}},\ \bibinfo {pages} {2758} (\bibinfo {year}
  {1995})}\BibitemShut {NoStop}%
\bibitem [{\citenamefont {Ueda}\ \emph {et~al.}(1996)\citenamefont {Ueda},
  \citenamefont {Kontani}, \citenamefont {Sigrist},\ and\ \citenamefont
  {Lee}}]{Ueda1996}%
  \BibitemOpen
  \bibfield  {author} {\bibinfo {author} {\bibfnamefont {K.}~\bibnamefont
  {Ueda}}, \bibinfo {author} {\bibfnamefont {H.}~\bibnamefont {Kontani}},
  \bibinfo {author} {\bibfnamefont {M.}~\bibnamefont {Sigrist}}, \ and\
  \bibinfo {author} {\bibfnamefont {P.~A.}\ \bibnamefont {Lee}},\ }\href
  {\doibase 10.1103/PhysRevLett.76.4650} {\bibfield  {journal} {\bibinfo
  {journal} {Phys. Rev. Lett.}\ }\textbf {\bibinfo {volume} {76}},\ \bibinfo
  {pages} {4650} (\bibinfo {year} {1996})}\BibitemShut {NoStop}%
\bibitem [{\citenamefont {Troyer}\ \emph {et~al.}(1996)\citenamefont {Troyer},
  \citenamefont {Kontani},\ and\ \citenamefont {Ueda}}]{Troyer1996}%
  \BibitemOpen
  \bibfield  {author} {\bibinfo {author} {\bibfnamefont {M.}~\bibnamefont
  {Troyer}}, \bibinfo {author} {\bibfnamefont {H.}~\bibnamefont {Kontani}}, \
  and\ \bibinfo {author} {\bibfnamefont {K.}~\bibnamefont {Ueda}},\ }\href
  {\doibase 10.1103/PhysRevLett.76.3822} {\bibfield  {journal} {\bibinfo
  {journal} {Phys. Rev. Lett.}\ }\textbf {\bibinfo {volume} {76}},\ \bibinfo
  {pages} {3822} (\bibinfo {year} {1996})}\BibitemShut {NoStop}%
\bibitem [{\citenamefont {Wei}\ \emph {et~al.}(2011)\citenamefont {Wei},
  \citenamefont {Qing-Zhen}, \citenamefont {Gen-Fu}, \citenamefont {Green},
  \citenamefont {Du-Ming}, \citenamefont {Jun-Bao},\ and\ \citenamefont
  {Yi-Ming}}]{Bao2011}%
  \BibitemOpen
  \bibfield  {author} {\bibinfo {author} {\bibfnamefont {B.}~\bibnamefont
  {Wei}}, \bibinfo {author} {\bibfnamefont {H.}~\bibnamefont {Qing-Zhen}},
  \bibinfo {author} {\bibfnamefont {C.}~\bibnamefont {Gen-Fu}}, \bibinfo
  {author} {\bibfnamefont {M.~A.}\ \bibnamefont {Green}}, \bibinfo {author}
  {\bibfnamefont {W.}~\bibnamefont {Du-Ming}}, \bibinfo {author} {\bibfnamefont
  {H.}~\bibnamefont {Jun-Bao}}, \ and\ \bibinfo {author} {\bibfnamefont
  {Q.}~\bibnamefont {Yi-Ming}},\ }\href
  {http://stacks.iop.org/0256-307X/28/i=8/a=086104} {\bibfield  {journal}
  {\bibinfo  {journal} {Chinese Physics Letters}\ }\textbf {\bibinfo {volume}
  {28}},\ \bibinfo {pages} {086104} (\bibinfo {year} {2011})}\BibitemShut
  {NoStop}%
\bibitem [{\citenamefont {Ye}\ \emph {et~al.}(2011)\citenamefont {Ye},
  \citenamefont {Chi}, \citenamefont {Bao}, \citenamefont {Wang}, \citenamefont
  {Ying}, \citenamefont {Chen}, \citenamefont {Wang}, \citenamefont {Dong},\
  and\ \citenamefont {Fang}}]{Ye2011}%
  \BibitemOpen
  \bibfield  {author} {\bibinfo {author} {\bibfnamefont {F.}~\bibnamefont
  {Ye}}, \bibinfo {author} {\bibfnamefont {S.}~\bibnamefont {Chi}}, \bibinfo
  {author} {\bibfnamefont {W.}~\bibnamefont {Bao}}, \bibinfo {author}
  {\bibfnamefont {X.~F.}\ \bibnamefont {Wang}}, \bibinfo {author}
  {\bibfnamefont {J.~J.}\ \bibnamefont {Ying}}, \bibinfo {author}
  {\bibfnamefont {X.~H.}\ \bibnamefont {Chen}}, \bibinfo {author}
  {\bibfnamefont {H.~D.}\ \bibnamefont {Wang}}, \bibinfo {author}
  {\bibfnamefont {C.~H.}\ \bibnamefont {Dong}}, \ and\ \bibinfo {author}
  {\bibfnamefont {M.}~\bibnamefont {Fang}},\ }\href {\doibase
  10.1103/PhysRevLett.107.137003} {\bibfield  {journal} {\bibinfo  {journal}
  {Phys. Rev. Lett.}\ }\textbf {\bibinfo {volume} {107}},\ \bibinfo {pages}
  {137003} (\bibinfo {year} {2011})}\BibitemShut {NoStop}%
\bibitem [{\citenamefont {Yan}\ \emph {et~al.}(2011)\citenamefont {Yan},
  \citenamefont {Gao}, \citenamefont {Lu},\ and\ \citenamefont
  {Xiang}}]{Yan2011}%
  \BibitemOpen
  \bibfield  {author} {\bibinfo {author} {\bibfnamefont {X.-W.}\ \bibnamefont
  {Yan}}, \bibinfo {author} {\bibfnamefont {M.}~\bibnamefont {Gao}}, \bibinfo
  {author} {\bibfnamefont {Z.-Y.}\ \bibnamefont {Lu}}, \ and\ \bibinfo {author}
  {\bibfnamefont {T.}~\bibnamefont {Xiang}},\ }\href {\doibase
  10.1103/PhysRevB.83.233205} {\bibfield  {journal} {\bibinfo  {journal} {Phys.
  Rev. B}\ }\textbf {\bibinfo {volume} {83}},\ \bibinfo {pages} {233205}
  (\bibinfo {year} {2011})}\BibitemShut {NoStop}%
\bibitem [{\citenamefont {Maiti}\ \emph {et~al.}(2011)\citenamefont {Maiti},
  \citenamefont {Korshunov}, \citenamefont {Maier}, \citenamefont
  {Hirschfeld},\ and\ \citenamefont {Chubukov}}]{Maiti2011}%
  \BibitemOpen
  \bibfield  {author} {\bibinfo {author} {\bibfnamefont {S.}~\bibnamefont
  {Maiti}}, \bibinfo {author} {\bibfnamefont {M.~M.}\ \bibnamefont
  {Korshunov}}, \bibinfo {author} {\bibfnamefont {T.~A.}\ \bibnamefont
  {Maier}}, \bibinfo {author} {\bibfnamefont {P.~J.}\ \bibnamefont
  {Hirschfeld}}, \ and\ \bibinfo {author} {\bibfnamefont {A.~V.}\ \bibnamefont
  {Chubukov}},\ }\href {\doibase 10.1103/PhysRevLett.107.147002} {\bibfield
  {journal} {\bibinfo  {journal} {Phys. Rev. Lett.}\ }\textbf {\bibinfo
  {volume} {107}},\ \bibinfo {pages} {147002} (\bibinfo {year}
  {2011})}\BibitemShut {NoStop}%
\bibitem [{\citenamefont {Xu}\ \emph {et~al.}(2013)\citenamefont {Xu},
  \citenamefont {Richard}, \citenamefont {Shi}, \citenamefont {van Roekeghem},
  \citenamefont {Qian}, \citenamefont {Razzoli}, \citenamefont {Rienks},
  \citenamefont {Chen}, \citenamefont {Ieki}, \citenamefont {Nakayama},
  \citenamefont {Sato}, \citenamefont {Takahashi}, \citenamefont {Shi},\ and\
  \citenamefont {Ding}}]{Xu2013}%
  \BibitemOpen
  \bibfield  {author} {\bibinfo {author} {\bibfnamefont {N.}~\bibnamefont
  {Xu}}, \bibinfo {author} {\bibfnamefont {P.}~\bibnamefont {Richard}},
  \bibinfo {author} {\bibfnamefont {X.}~\bibnamefont {Shi}}, \bibinfo {author}
  {\bibfnamefont {A.}~\bibnamefont {van Roekeghem}}, \bibinfo {author}
  {\bibfnamefont {T.}~\bibnamefont {Qian}}, \bibinfo {author} {\bibfnamefont
  {E.}~\bibnamefont {Razzoli}}, \bibinfo {author} {\bibfnamefont
  {E.}~\bibnamefont {Rienks}}, \bibinfo {author} {\bibfnamefont {G.-F.}\
  \bibnamefont {Chen}}, \bibinfo {author} {\bibfnamefont {E.}~\bibnamefont
  {Ieki}}, \bibinfo {author} {\bibfnamefont {K.}~\bibnamefont {Nakayama}},
  \bibinfo {author} {\bibfnamefont {T.}~\bibnamefont {Sato}}, \bibinfo {author}
  {\bibfnamefont {T.}~\bibnamefont {Takahashi}}, \bibinfo {author}
  {\bibfnamefont {M.}~\bibnamefont {Shi}}, \ and\ \bibinfo {author}
  {\bibfnamefont {H.}~\bibnamefont {Ding}},\ }\href {\doibase
  10.1103/PhysRevB.88.220508} {\bibfield  {journal} {\bibinfo  {journal} {Phys.
  Rev. B}\ }\textbf {\bibinfo {volume} {88}},\ \bibinfo {pages} {220508}
  (\bibinfo {year} {2013})}\BibitemShut {NoStop}%
\bibitem [{\citenamefont {Yanagi}\ and\ \citenamefont
  {Ueda}(2014)}]{Yanagi2014}%
  \BibitemOpen
  \bibfield  {author} {\bibinfo {author} {\bibfnamefont {Y.}~\bibnamefont
  {Yanagi}}\ and\ \bibinfo {author} {\bibfnamefont {K.}~\bibnamefont {Ueda}},\
  }\href {\doibase 10.1103/PhysRevB.90.085113} {\bibfield  {journal} {\bibinfo
  {journal} {Phys. Rev. B}\ }\textbf {\bibinfo {volume} {90}},\ \bibinfo
  {pages} {085113} (\bibinfo {year} {2014})}\BibitemShut {NoStop}%
\bibitem [{\citenamefont {Khatami}\ \emph {et~al.}(2014)\citenamefont
  {Khatami}, \citenamefont {Singh}, \citenamefont {Pickett},\ and\
  \citenamefont {Scalettar}}]{Khatami2014}%
  \BibitemOpen
  \bibfield  {author} {\bibinfo {author} {\bibfnamefont {E.}~\bibnamefont
  {Khatami}}, \bibinfo {author} {\bibfnamefont {R.~R.~P.}\ \bibnamefont
  {Singh}}, \bibinfo {author} {\bibfnamefont {W.~E.}\ \bibnamefont {Pickett}},
  \ and\ \bibinfo {author} {\bibfnamefont {R.~T.}\ \bibnamefont {Scalettar}},\
  }\href {\doibase 10.1103/PhysRevLett.113.106402} {\bibfield  {journal}
  {\bibinfo  {journal} {Phys. Rev. Lett.}\ }\textbf {\bibinfo {volume} {113}},\
  \bibinfo {pages} {106402} (\bibinfo {year} {2014})}\BibitemShut {NoStop}%
\bibitem [{\citenamefont {Yamada}(2014)}]{Yamada2014}%
  \BibitemOpen
  \bibfield  {author} {\bibinfo {author} {\bibfnamefont {A.}~\bibnamefont
  {Yamada}},\ }\href {\doibase arXiv:1408.4879} {\bibfield  {journal} {\bibinfo
   {journal} {arXiv}\ }\textbf {\bibinfo {volume} {1408}},\ \bibinfo {pages}
  {4879} (\bibinfo {year} {2014})}\BibitemShut {NoStop}%
\bibitem [{\citenamefont {Tsai}\ and\ \citenamefont
  {Kivelson}(2006)}]{Tsai2006}%
  \BibitemOpen
  \bibfield  {author} {\bibinfo {author} {\bibfnamefont {W.-F.}\ \bibnamefont
  {Tsai}}\ and\ \bibinfo {author} {\bibfnamefont {S.~A.}\ \bibnamefont
  {Kivelson}},\ }\href {\doibase 10.1103/PhysRevB.73.214510} {\bibfield
  {journal} {\bibinfo  {journal} {Phys. Rev. B}\ }\textbf {\bibinfo {volume}
  {73}},\ \bibinfo {pages} {214510} (\bibinfo {year} {2006})}\BibitemShut
  {NoStop}%
\bibitem [{\citenamefont {Wenzel}\ and\ \citenamefont
  {Janke}(2009)}]{Wenzel2009}%
  \BibitemOpen
  \bibfield  {author} {\bibinfo {author} {\bibfnamefont {S.}~\bibnamefont
  {Wenzel}}\ and\ \bibinfo {author} {\bibfnamefont {W.}~\bibnamefont {Janke}},\
  }\href {\doibase 10.1103/PhysRevB.79.014410} {\bibfield  {journal} {\bibinfo
  {journal} {Phys. Rev. B}\ }\textbf {\bibinfo {volume} {79}},\ \bibinfo
  {pages} {014410} (\bibinfo {year} {2009})}\BibitemShut {NoStop}%
\bibitem [{\citenamefont {Chakraborty}\ \emph {et~al.}(2011)\citenamefont
  {Chakraborty}, \citenamefont {S\'en\'echal},\ and\ \citenamefont
  {Tremblay}}]{Chakraborty2011}%
  \BibitemOpen
  \bibfield  {author} {\bibinfo {author} {\bibfnamefont {S.}~\bibnamefont
  {Chakraborty}}, \bibinfo {author} {\bibfnamefont {D.}~\bibnamefont
  {S\'en\'echal}}, \ and\ \bibinfo {author} {\bibfnamefont {A.-M.~S.}\
  \bibnamefont {Tremblay}},\ }\href {\doibase 10.1103/PhysRevB.84.054545}
  {\bibfield  {journal} {\bibinfo  {journal} {Phys. Rev. B}\ }\textbf {\bibinfo
  {volume} {84}},\ \bibinfo {pages} {054545} (\bibinfo {year}
  {2011})}\BibitemShut {NoStop}%
\bibitem [{\citenamefont {Ying}\ \emph {et~al.}(2014)\citenamefont {Ying},
  \citenamefont {Mondaini}, \citenamefont {Sun}, \citenamefont {Paiva},
  \citenamefont {Fye},\ and\ \citenamefont {Scalettar}}]{Ying2014}%
  \BibitemOpen
  \bibfield  {author} {\bibinfo {author} {\bibfnamefont {T.}~\bibnamefont
  {Ying}}, \bibinfo {author} {\bibfnamefont {R.}~\bibnamefont {Mondaini}},
  \bibinfo {author} {\bibfnamefont {X.~D.}\ \bibnamefont {Sun}}, \bibinfo
  {author} {\bibfnamefont {T.}~\bibnamefont {Paiva}}, \bibinfo {author}
  {\bibfnamefont {R.~M.}\ \bibnamefont {Fye}}, \ and\ \bibinfo {author}
  {\bibfnamefont {R.~T.}\ \bibnamefont {Scalettar}},\ }\href {\doibase
  10.1103/PhysRevB.90.075121} {\bibfield  {journal} {\bibinfo  {journal} {Phys.
  Rev. B}\ }\textbf {\bibinfo {volume} {90}},\ \bibinfo {pages} {075121}
  (\bibinfo {year} {2014})}\BibitemShut {NoStop}%
\bibitem [{\citenamefont {Hassan}\ and\ \citenamefont
  {S\'en\'echal}(2013)}]{Hassan2013}%
  \BibitemOpen
  \bibfield  {author} {\bibinfo {author} {\bibfnamefont {S.}~\bibnamefont
  {Hassan}}\ and\ \bibinfo {author} {\bibfnamefont {D.}~\bibnamefont
  {S\'en\'echal}},\ }\href {\doibase 10.1103/PhysRevLett.110.096402} {\bibfield
   {journal} {\bibinfo  {journal} {Phys. Rev. Lett.}\ }\textbf {\bibinfo
  {volume} {110}},\ \bibinfo {pages} {096402} (\bibinfo {year}
  {2013})}\BibitemShut {NoStop}%
\bibitem [{\citenamefont {Schwandt}\ \emph {et~al.}(2009)\citenamefont
  {Schwandt}, \citenamefont {Alet},\ and\ \citenamefont
  {Capponi}}]{Schwandt2009}%
  \BibitemOpen
  \bibfield  {author} {\bibinfo {author} {\bibfnamefont {D.}~\bibnamefont
  {Schwandt}}, \bibinfo {author} {\bibfnamefont {F.}~\bibnamefont {Alet}}, \
  and\ \bibinfo {author} {\bibfnamefont {S.}~\bibnamefont {Capponi}},\ }\href
  {\doibase 10.1103/PhysRevLett.103.170501} {\bibfield  {journal} {\bibinfo
  {journal} {Phys. Rev. Lett.}\ }\textbf {\bibinfo {volume} {103}},\ \bibinfo
  {pages} {170501} (\bibinfo {year} {2009})}\BibitemShut {NoStop}%
\bibitem [{\citenamefont {Sakai}\ \emph {et~al.}(2012)\citenamefont {Sakai},
  \citenamefont {Sangiovanni}, \citenamefont {Civelli}, \citenamefont {Motome},
  \citenamefont {Held},\ and\ \citenamefont {Imada}}]{Sakai2012}%
  \BibitemOpen
  \bibfield  {author} {\bibinfo {author} {\bibfnamefont {S.}~\bibnamefont
  {Sakai}}, \bibinfo {author} {\bibfnamefont {G.}~\bibnamefont {Sangiovanni}},
  \bibinfo {author} {\bibfnamefont {M.}~\bibnamefont {Civelli}}, \bibinfo
  {author} {\bibfnamefont {Y.}~\bibnamefont {Motome}}, \bibinfo {author}
  {\bibfnamefont {K.}~\bibnamefont {Held}}, \ and\ \bibinfo {author}
  {\bibfnamefont {M.}~\bibnamefont {Imada}},\ }\href {\doibase
  10.1103/PhysRevB.85.035102} {\bibfield  {journal} {\bibinfo  {journal} {Phys.
  Rev. B}\ }\textbf {\bibinfo {volume} {85}},\ \bibinfo {pages} {035102}
  (\bibinfo {year} {2012})}\BibitemShut {NoStop}%
\bibitem [{\citenamefont {Chen}\ \emph {et~al.}(2012)\citenamefont {Chen},
  \citenamefont {Meng}, \citenamefont {Pruschke}, \citenamefont {Moreno},\ and\
  \citenamefont {Jarrell}}]{Chen2012}%
  \BibitemOpen
  \bibfield  {author} {\bibinfo {author} {\bibfnamefont {K.-S.}\ \bibnamefont
  {Chen}}, \bibinfo {author} {\bibfnamefont {Z.~Y.}\ \bibnamefont {Meng}},
  \bibinfo {author} {\bibfnamefont {T.}~\bibnamefont {Pruschke}}, \bibinfo
  {author} {\bibfnamefont {J.}~\bibnamefont {Moreno}}, \ and\ \bibinfo {author}
  {\bibfnamefont {M.}~\bibnamefont {Jarrell}},\ }\href {\doibase
  10.1103/PhysRevB.86.165136} {\bibfield  {journal} {\bibinfo  {journal} {Phys.
  Rev. B}\ }\textbf {\bibinfo {volume} {86}},\ \bibinfo {pages} {165136}
  (\bibinfo {year} {2012})}\BibitemShut {NoStop}%
\bibitem [{\citenamefont {Lifshitz}(1960)}]{Lifshitz1960}%
  \BibitemOpen
  \bibfield  {author} {\bibinfo {author} {\bibfnamefont {I.~M.}\ \bibnamefont
  {Lifshitz}},\ }\href@noop {} {\bibfield  {journal} {\bibinfo  {journal} {Zh.
  Eksp. Teor. Fiz.}\ }\textbf {\bibinfo {volume} {38}},\ \bibinfo {pages}
  {1569} (\bibinfo {year} {1960})}\BibitemShut {NoStop}%
\bibitem [{\citenamefont {L\"auchli}\ \emph {et~al.}(2002)\citenamefont
  {L\"auchli}, \citenamefont {Wessel},\ and\ \citenamefont
  {Sigrist}}]{Lauchli2002}%
  \BibitemOpen
  \bibfield  {author} {\bibinfo {author} {\bibfnamefont {A.}~\bibnamefont
  {L\"auchli}}, \bibinfo {author} {\bibfnamefont {S.}~\bibnamefont {Wessel}}, \
  and\ \bibinfo {author} {\bibfnamefont {M.}~\bibnamefont {Sigrist}},\ }\href
  {\doibase 10.1103/PhysRevB.66.014401} {\bibfield  {journal} {\bibinfo
  {journal} {Phys. Rev. B}\ }\textbf {\bibinfo {volume} {66}},\ \bibinfo
  {pages} {014401} (\bibinfo {year} {2002})}\BibitemShut {NoStop}%
\bibitem [{\citenamefont {Scalapino}\ and\ \citenamefont
  {Trugman}(1996)}]{Scalapino1996}%
  \BibitemOpen
  \bibfield  {author} {\bibinfo {author} {\bibfnamefont {D.}~\bibnamefont
  {Scalapino}}\ and\ \bibinfo {author} {\bibfnamefont {S.}~\bibnamefont
  {Trugman}},\ }\href {\doibase 10.1080/01418639608240361} {\bibfield
  {journal} {\bibinfo  {journal} {Philos. Mag. B}\ }\textbf {\bibinfo {volume}
  {74}},\ \bibinfo {pages} {607} (\bibinfo {year} {1996})}\BibitemShut
  {NoStop}%
\bibitem [{\citenamefont {Sandvik}(1997)}]{Sandvik1997}%
  \BibitemOpen
  \bibfield  {author} {\bibinfo {author} {\bibfnamefont {A.~W.}\ \bibnamefont
  {Sandvik}},\ }\href {\doibase 10.1103/PhysRevB.56.11678} {\bibfield
  {journal} {\bibinfo  {journal} {Phys. Rev. B}\ }\textbf {\bibinfo {volume}
  {56}},\ \bibinfo {pages} {11678} (\bibinfo {year} {1997})}\BibitemShut
  {NoStop}%
\end{thebibliography}%

\end{document}